# Low voltage and time constant organic synapse-transistor

*Simon Desbief[1], Adrica Kyndiah[2,3], David Guerin[1], Denis Gentili,[2] Mauro Murgia[2], Stéphane Lenfant[1], Fabien Alibart[1,] Tobias Cramer[2,5], Fabio Biscarini[4], and Dominique Vuillaume[1,a]*.

1) Institute for Electronics Microelectronics and Nanotechnology, CNRS and University of Lille, Av. Poincaré, F-59652cedex, Villeneuve d'Ascq, France.

2) Consiglio Nazionale delle Ricerche, Istituto per lo Studio dei Materiali Nanostrutturati (CNR-ISMN) Via P. Gobetti 101, 40129 Bologna, Italy.

3) Alma Mater Studiorum-Università degli Studi di Bologna, Dipartimento di Chimica "G. Ciamician", Via Selmi 2, 40127, Bologna, Italy

4) Life Science Dept., Univ. Modena and Reggio Emilia, Via Campi 183, 41125 Modena, Italy.

5) Università di Bologna, Dipartimento di Fisica e Astronomia, Viale Berti Pichat 6/2, Bologna, Italy.

a) Corresponding author : dominique.vuillaume@iemn.univ-lille1.fr

## Abstract.

We report on an artificial synapse, an organic synapse-transistor (synapstor) working at 1 volt and with a typical response time in the range 100-200 ms. This device (also called NOMFET, Nanoparticle Organic Memory Field Effect Transistor) combines a memory and a transistor effect in a single device. We demonstrate that short-term plasticity (STP), a typical synaptic behavior, is observed when stimulating the device with input spikes of 1 volt. Both significant facilitating and depressing behaviors of this artificial synapse are observed with a relative amplitude of about 50% and a dynamic response < 200 ms. From a series of *in-situ* experiments, i.e. measuring the current-voltage characteristic curves *in-situ* and in real time, during the growth of the pentacene over a network of gold nanoparticles, we elucidate these results by analyzing the relationship between the organic film morphology and the transport properties. This synapstor works at a low energy of about 2 nJ/spike. We discuss the implications of these results for the development of neuro-inspired computing architectures and interfacing with biological neurons.

# 1. Introduction.

There is a growing interest in neuro-inspired devices based on emerging technologies beyond the existing silicon CMOS implementation of artificial neural networks (e.g. see recent reviews in a special issue[1] and in Refs. [2,3]). Organic bioelectronics is emerging as a viable technological approach aiming at interfacing organic devices with cells and neurons.[4-6] We have recently demonstrated how we can use charge trapping/detrapping in an array of gold nanoparticles (NPs) at the $SiO_2$/pentacene interface to design a synapse-transistor (synapstor) mimicking the dynamic plasticity of a biological synapse.[7,8] This device called NOMFET (Nanoparticle Organic Memory Field Effect Transistor) combines a memory and a transistor effect in a single device. NOMFETs have been initially designed for neuro-inspired computing architectures (artificial neural networks). This device (which is memristor-like)[8] mimics short-term plasticity (STP)[7] and temporal correlation plasticity (STDP, spike-timing dependent plasticity)[8] of biological spiking synapses, two "functions" at the basis of learning processes. A compact model was developed[9] and we demonstrated an associative memory, which can be trained to exhibit a Pavlovian response.[10] However, these devices were limited to work with spikes in the range of few tens of volts and time scale of 1-100s. Both fields (neuro-inspired computing and bioelectronics) require devices working at lower voltages (to save energy consumption during computing and because action potentials in synapse and neurons have amplitudes of around 100 mV) and higher speed (e.g. synapse and neurons work at around kHz).

Here, we report on a synapstor working at 1 volt and with a typical response time constant in the range 100-200 ms. To gain a better insight of the NOMFET behavior, we performed a series of *in-situ* experiments where the response of the device is studied during the growth of the organic semiconductor thin film. We analyzed the relationship between the organic film morphology and the transport properties by measuring the current-voltage characteristic curves, *in-situ* and in real time, during the growth of the pentacene over the NP network. Second, we performed temperature-dependent measurements of the charge carrier mobility to assess the effect of NPs on charge transport properties. Finally, based on the conclusions drawn from these experiments, we report on an optimized NOMFET fabrication process getting a larger hole mobility of the NOMFET (~ 0.1 cm$^2$/Vs, instead of 10$^{-3}$ - 10$^{-2}$ cm$^2$/Vs in our previous work)[7] and a faster response time constant (hundreds of ms instead of few seconds).

## 2. Materials and Methods

*2.1. Si/SiO$_2$ wafer substrate and source/drain electrodes.*

The NOMFETs were processed using a bottom-gate electrode configuration. We used highly-doped (resistivity ~10$^{-3}$ Ω.cm) n-type silicon covered with a thermally grown 200 nm thick silicon dioxide (grown at 1100°C during 135 min in a dry oxygen flow (2L/min) and followed by a post-oxidation annealing at 900°C during 30 min under a nitrogen flow (2L/min) in order to reduce the presence of defects into the oxide). Electrodes (titanium/gold (20/200 nm)) were deposited on the substrate by vacuum evaporation and patterned by e-beam lithography for linear source and drain gold electrodes (length L=1 to 50 μm and width W=1000 μm) and by optical lithography for interdigitated source and drain gold electrodes (length L= 20 and 40 μm, width W=11200 and 22400 μm). Before use, these wafers were cleaned following the protocol : i) sonication in acetone for 15 min and isopropanol (15 min), dried in N$_2$. ii) Piranha solution

($H_2SO_4$ /$H_2O_2$, 2/1 v/v) for 15 minutes (*Caution: preparation of the piranha solution is highly exothermic and reacts violently with organics*). iii) Rinse with DI water, copiously. iv) Etch with HF 2% for 5s, rinse with DI water, copiously, dry in $N_2$ and bake at 120° (hot plate), or clean by ultraviolet irradiation in an ozone atmosphere (ozonolysis) for 30 minutes. The cleaned substrates are used immediately (see surface functionalization below).

### *2.2. Au NPs synthesis.*

Colloidal solutions of citrate-capped Au NP (10 ± 3 nm in diameter) were synthesis as follows.[11] Charge stabilized Au nanoparticles were synthesized by the reduction of chloroauric acid in water. To obtain a 100 mL aqueous solution of Au nanoparticles, a solution with 1 mL of $HAuCl_4 4H_2O$ (1% w/v) in 79 mL of $H_2O$ was first prepared. A 20 mL reducing solution with 4 mL of trisodium citrate (1% w/v) and 80 $\mu$L of tannic acid (1% w/v) in 16 mL of $H_2O$ was then added rapidly to the Au solution (all solutions at 60 °C). The mixed solution was boiled for 10 min before being cooled down to room temperature. A continuous stirring was applied throughout the process. The resulting reddish solution contained typically 10nm Au. The NP size (10 ± 3 nm in diameter) and density is calculated from SEM images of NPs arrays on the surface (as the one in Fig. S1 in SI).

### *2.3. $SiO_2$ surface functionalization and NP deposition.*

Immediately after the cleaning (see above), the $SiO_2$ gate dielectric was functionalized by self-assembled monolayer (SAM) to anchor gold nanoparticles (NPs) on the surface.[12] The $SiO_2$ surface was functionalized by immersion in a solution of (3-aminopropyl)-trimethoxysilane (APTMS) molecules diluted in methanol at a concentration 1$\mu$L/mL for 1h.[13] The reaction took place in a glove-box with a controlled atmosphere (nitrogen, with less than 1 ppm of oxygen and water vapor). Excess, non-reacted, molecules were removed by rinse in methanol under sonication. This sample was subsequently dried under nitrogen stream. Static water contact angle was 19°, a common value for hydrophilic NH2-terminated surfaces.[13] This sample was then immediately dipped in a colloidal solution of citrate-stabilized Au-NP for 24h. This procedure yields an array of NPs with a density of about $4-5 \times 10^{10}$ NP/cm$^2$ (see details in the Supporting Information, Fig. S1). The sample was then cleaned with deionized water and isopropanol, and dried under nitrogen stream. The NPs deposited on SiO$_2$ do not form a continuous film or large clusters, rather they are adsorbed as individual entities. They do not coalesce (mainly due to coulombic repulsion between the negatively charged citrate layer), and they exhibit a characteristic length scales, i.e. a density, that we can adjust with the concentration of the Au NPs in the solution and the time of deposition.[7] Indeed, we have previously demonstrated that an optimized density of NPs for the NOMFETs is around $5 \times 10^{10}$ NP/cm$^2$.[7]

### 2.4. OTS functionalization.

Octadecyltricholorosilane (OTS) molecules were used as follow.[14, 15] The silanization reaction was carried out in a glovebox under nitrogen atmosphere but non-anhydrous solvents were used to favor hydrolysis of -SiCl3 functions. The freshly cleaned substrate was immersed for 2h in a $10^{-3}$ M solution of OTS in a mixture of n-hexane and dichloromethane (70:30 v/v). The device was rinsed thoroughly by sonication in dichloromethane (2 times) then blown with dry nitrogen.

*2.5. Pentacene evaporation.*

For *in-situ* experiments. Pentacene was evaporated at a rate of 0.5 ML/min (~0.125 Å/s, 1 ML of pentacene ≈ 15 Å). The substrate was kept at room temperature. About 20 ML (30 nm) were evaporated. For device measurements (*ex-situ*), a 35 nm thick pentacene film was evaporated at a rate of 0.1 Å/s. The substrate was kept at 60°C. 4 devices were fabricated in parallel during the same pentacene evaporation run: i)A reference pentacene transistor without NPs (and without APTMS SAM on $SiO_2$) - referred to as P5. ii) A second pentacene transistor with an OTS functionalized $SiO_2$ gate dielectric - referred to as OTS-P5. iii) A "standard" NOMFET, with APTMS and NPs - referred to as APTMS-NP-P5. iv) The "optimized" NOMFET with the OTS treatment and referred to as APTMS-NP-OTS-P5.

*2.6. AFM measurements.*

The surface morphology of the organic layer was determined by imaging with a Bruker ICON atomic force microscope (AFM) in tapping mode. Silicon cantilevers from Bruker (NCHV model) were used to acquire AFM images of 5 x 5 $\mu m^2$ at 1Hz with a resolution of 512 x 512 pixels.

*2.7. Contact angle measurements.*

The wettability of the surface at different stages of the fabrication was assessed by measuring the water contact angle. We measured the static water contact angle with a remote-computer controlled goniometer system (DIGIDROP by GBX, France). We deposited a drop (10-30 μL) of deionized water (18 MΩ.cm-1) on the surface, and the projected image was acquired and stored by the computer. Contact angles were extracted by contrast contour image analysis software.

These angles were determined a few seconds after application of the drop. These measurements were carried out in a clean room (ISO 6) where the relative humidity (50%) and the temperature (22 °C) are controlled. The precision with these measurements are ±2°.

*2.8. Electrical measurements.*

For *in-situ* measurements, just before pentacene evaporation, samples were transferred in the home-built vacuum evaporation chamber,[16] the bottom gate and the bottom source/drain electrodes were electrically connected, and drain current-gate voltage ($I_D$-$V_G$) curves were recorded *in-situ* from $V_G$= 30V to -30V both in linear ($V_D$=-1V) and saturation regime ($V_D$=-30V) as pentacene was sublimed at a rate of 0.5 ML/min (substrate at room-temperature, 1 ML of pentacene ≈1.5 nm). When pentacene thickness reached about 20 MLs, deposition was stopped and temperature dependent measurements were carried out by recording $I_D$-$V_G$ scans every 30 s while the temperature was varied from -160°C to RT. Hole mobility in linear and saturation regimes was extracted as usual from standard transistor equations. A low-current source-measure unit (SMU) Keithley 6430 is used to apply the gate–source voltage ($V_{GS}$) and to measure the gate current ($I_G$), and a dual SMU Keithley 2612 is used to apply the drain–source voltages ($V_{DS}$) and to measure the respective drain–source currents ($I_{DS}$).

The NOMFET electrical *ex-situ* characteristics were measured with an Agilent 4155C semiconductor parameter analyzer, the input pulses were delivered by a pulse generator (Tabor 5061 or Keithley 2636A). The electrodes of the NOMFET were contacted with a shielded micro-manipulator probe station (Suss Microtec) in the dark.

## 3. Results and discussion.

### *3.1. In-situ experiments.*

Figure 1 shows the evolution of the hole mobility (in saturation regime) as a function of the thickness of the deposited pentacene film. While mobility in linear and saturation regimes are slightly different ($\mu_{sat} > \mu_{lin}$ by a factor of about 1.5-2), their evolution with pentacene thickness is the same. The *in-situ* measurements allow us to evidence the formation of the channel layer and charge transport (holes) along the dielectric interface functionalized with NPs, as the control in thickness is a fraction of monolayer. First we investigate pentacene deposition and channel formation on the NP functionalized substrates. Fig. 1 shows how the mobility of this transistor evolves as a function of the nominal pentacene layer thickness $\Theta$ (expressed in monolayers ML by taking into account 1ML of pentacene $\approx$ 1.5nm) during the *in-situ* experiment. Below a critical thickness of $\Theta_{C1,NP}$ = 5.7 ML (square symbol, NOMFET without OTS treatment) no transistor current can be measured and the mobility cannot be extracted. At $\Theta_{C1,NP}$ percolation of conducting islands sets in and a first continuous pathway connects source and drain electrode. We measure current modulation in the transfer characteristics that allows us to extract the mobility $\mu$ as a function of $\Theta \geq \Theta_{C1,NP}$. Upon addition of pentacene the transistor current and the mobility both increase rapidly over orders of magnitude. The increase does not follow a single power law but contains an inflection point at $\Theta_{C2,NP}$ = 8.4 ML beyond which a second phase of steep increase in mobility is observed. In Fig 1, both regions are fitted by a power law $\sim(\Theta-\Theta_C)^\gamma$, with the critical coverage $\Theta_C$ indicating the creation of a percolation path between the electrodes, and the critical exponent $\gamma$ related to the lattice geometry and dimensionality of the underlying conducting network.[17] From the fit we find an increase in exponent from 1.1 in the first phase to

1.25 in the second phase of mobility increase, demonstrating a change in growth mode at $\Theta = 8.4$ ML (i.e. about 12 nm) caused by the presence of NPs (diameter of about 10 nm). Finally, saturation of the curve is slowly attained at $\Theta \approx 20$ ML of deposited pentacene to yield a saturated charge mobility of $\mu = 3 \times 10^{-3}$ cm$^2$/Vs. This mobility value is on agreement with the one we previously measured on NOMFET.[7] The observed evolution of the mobility with $\Theta$ clearly results from the more disordered organisation of the semiconducting film grown on the NPs. In pentacene transistors grown on a smooth dielectric interface, percolation is observed before the completion of the first monolayer ($\Theta_{C,2D} \sim 0.7$ ML) following the theory of 2D percolation[18, 19] (see Fig. S2, Supplementary Information for in situ $\mu$ vs $\Theta$ measurements without NPs). Here, the shift of the percolation threshold above 5 MLs demonstrates a transition of pentacene growth to a 3D growth mode induced by the NPs adsorbed on the surface. Further complexity in the formation of additional transport paths leads to the formation of the inflection point. Only when the nominal thickness exceeds 8 ML a second critical regime emerges where an additional steep increase of mobility is observed. This second threshold happens at a thickness which exceeds the diameter of the citrate capped NPs (about 10 nm). This finding indicates that efficient transport paths are formed when the NPs start to get covered by the pentacene layer. Therefore, we infer that a more disordered pentacene film is obtained with NPs, as also previously revealed by AFM images with smaller pentacene grain size for the NOMFET compare to pentacene transistor without NPs (see Fig. 2 in Ref. [7] and more images in the next section).

Complementary to these measurements we performed temperature dependent characterization of the transistor once the final layer thickness of 20 ML is achieved. Fig. 2 depicts the mobility as a function of inverse temperature. The Arrhenius plot shows strongly temperature activated behavior which can be described by a single effective barrier of $E_A = 123$ meV. We assign this barrier to a trap state in the HOMO-LUMO gap, near the HOMO.[20-22] Extrapolation to infinite temperature yields a trap-free mobility of $\mu_0 = 0.26$ cm$^2$/Vs. Without the NPs, we measured an activation energy $E_A = 65$ meV (see Fig. S3 in the supporting information). This latter value is in agreement with previous reports for pentacene OFET (in the range 20-80 meV).[20, 23-26] Our findings point to two possible reasons for the weak performance of the NOMFET grown on citrate capped NPs. First, the pentacene morphology is disordered due to the presence of NPs, and only transport paths spanning at a distance from the dielectric interface can evolve. The electronic coupling and charge mobility are reduced along such a path. The extrapolated trap free mobility is an order of magnitude lower than typical values found in unperturbed pentacene transistors.[27] Second, deep trap states further reduce mobility and lead to activated transport behavior with higher energy barrier. In the NOMFETs, once a sufficiently thick pentacene film is deposited (above about 10 ML, Fig. 1), each nanoparticle is surrounded by pentacene, thus each NP acts as a "shallow trap" for charge carriers moving across the pentacene thin film, explaining the increased activation energy $E_A$ with the NPs.

Since low charge carrier mobility in NOMFET is the main cause of the slow dynamics of its synaptic behavior,[7, 8] we modify the fabrication protocol adding a silanization treatment after the NPs deposition. The findings from the in-situ experiment prompted this strategy. Reduced mobility was already observed in the APTMS modified dielectric without the presence of NPs.[28] Therefore, both the surface chemistry that exposes the highly polarizable groups and the morphological changes of the semiconductor as induced by the presence of NPs (given that a sufficiently thick pentacene layer is deposited), contribute to the decrease of charge transfer velocity. Surface functionalization by molecules such as octadecyltrichlorosilane (OTS) is known to improve mobility in OFET as they bind to polar groups at the surface and form an oriented monolayer which exposes non-polar alkyl chains, thereby decreasing the surface energy.[29-31] We treated the NPs/APTMS substrates in a solution of OTS (see details in Materials and Methods) according to an established protocol.[14, 15, 32-34] Chlorosilane head groups of the OTS molecules are prone to react with both the terminal amine group of APTMS and the -OH and COO⁻ groups of citrate capping the NPs.[35-37] Moreover, OTS can polymerized laterally, though siloxane bonds, forming a capping layer over the surface and NPs. Evidence of the correct grafting of OTS is the observation of an increase of the water contact angle from ~30-40° for APTMS/NPs surface to ~90-100° after silanization. This latter value is consistent with 100-110° reported for OTS monolayer on a smooth surface (SiO$_2$),[34] (we observe contact angles of 106-109° for APTMS/OTS bilayer on flat SiO$_2$).

Now we compare the above results with *in-situ* experiments performed after passivation of the NP surface with OTS. The increase of mobility as a function of $\Theta$, as shown in Fig.1 (red dots), clearly confirms the improved performance. Percolation sets in earlier at $\Theta_{C1,NP/OTS}$ = 5.2 ML, but remains still beyond the sub-monolayer regime. Here the percolation process follows a simpler pattern that is described by a single power-law in the initial critical regime. The final increase in mobility saturates slowly and only at $\Theta$ = 18 ML a stable value of $\mu$ = 3 x 10$^{-2}$ cm$^2$/Vs is obtained which exceeds by more than an order of magnitude the non-OTS passivated device. Temperature dependent measurements demonstrate also for the OTS passivated NOMFET a thermally activated transport, but the activation barrier is clearly reduced to $E_A$ = 71 meV, close to the one for pentacene transistor (see Fig. S2, Supporting Information). Extrapolation to infinite temperature results in $\mu_0$ = 0.25 cm$^2$/Vs, as for the non-OTS passivated device. The comparison of the two types of NOMFETs provides a clear interpretation of the improvements in electrical performance (see section below). Most important, we achieved a smoothening of the trap states that limited transport in the non-OTS passivated device. The exposure of the polar citric acid groups at the NP surface leads to strong electrostatic interactions with carriers and consequently trapping. OTS reacts directly with these groups[36, 37] making the NP surface hydrophobic by exposing the alkyl chain. As a consequence the mobility rises by more than one order of magnitude when operated at room temperature (see below). However, both kind of surfaces demonstrate critical limitations of transport when compared to transistors grown on smooth dielectric surfaces: the NPs perturb the morphology in the first monolayers and give rise to a hindered transport path with reduced electronic coupling. This finding is independent of the NP functionalization and in both investigated cases we find maximum "trap-free" mobilities $\mu_0 = 0.25$ cm$^2$/Vs by extrapolation.

*3.2. Synaptic performances of the NOMFETs.*

In parallel with these *in-situ* experiments, we fabricated NOMFET with and without the OTS treatments to assess improvements of their synaptic behavior (devices referred to as "APTMS-NP-P5" and "APTMS-NP-OTS-P5", respectively). The NP deposition and OTS protocols were the same as for the *in-situ* measurement, then a 30 nm thick pentacene was directly evaporated to complete the device (see Materials and Methods). Fig. 3 shows typical $I_D$-$V_G$ measurements in saturation and the $I_D$-$V_D$ characteristics. For comparison, we also fabricated transistors without the NPs during the same run of pentacene deposition (pentacene directly on SiO$_2$, referred to as "P5" and pentacene on OTS functionalized SiO$_2$, referred to as "OTS-P5"). Fig. 4 shows the extracted mobilities for the four types of device. For the OTS-treated NOMFET ("APTMS-NP-OTS-P5"), we obtained a mobility of ~ 0.1 cm$^2$/Vs whatever the channel length in the 1-50 μm range, on a part with the mobility of pentacene OFET (without OTS, Fig. 4 device "P5"). Albeit smaller than the one for OTS-treated pentacene OFET (0.4-0.5 cm$^2$/Vs, Fig. 4, "OTS-P5"), this value is strongly improved compared to standard NOMFETs (without OTS treatment), which is around 10$^{-3}$ cm$^2$/Vs (Fig. 4, "APTMS-NP-P5") consistent with our previous results.[7, 8] AFM images of the pentacene (Fig. 5) clearly shows that the OTS-treatment increases the grain size of the pentacene film of the NOMFETs, as already reported for pentacene OFET.[29, 38] To measure the synaptic properties, the NOMFET is used as a pseudo two-terminal device (inset Figure 6): the source (S) and gate (G) electrodes are connected together and used as the input terminal of the device, and the drain (D) is used as the output terminal (virtually grounded through the amperometer). A sequence of spikes with amplitude $V_{spike}$ and at different frequencies is applied at the G/S input and we measure the output drain current. Figure 6-a shows a typical result for the optimized NOMFET (APTMS-NP-OTS-P5) developed in this work. As in previous reports,[7, 8] we clearly observe the STP (short term plasticity) behavior, mimicking a biological synapse.[39] In brief, the output current (or equivalently the NOMFET conductance) shows a depressing behavior (decrease of the current with the number of spikes) for the spike at the highest frequencies and a facilitating behavior (increase of the current) for the lowest ones, similar to the synaptic weight (the signal transmittance through the synapse) of a biological synapse. The major improvements, compared to our previous results (Fig. 6-b), are twofold: i) STP behavior is obtained with spikes of amplitude $V_{spike}$ = -1V instead of -10 to -20V (for a channel length L=1 μm) in Ref.[7]; ii) the characteristic response time constant is lowered to hundreds of ms, while we measured few seconds (at same channel length L = 1 μm).[7] This time constant is obtained by fitting (red squares Fig. 6) an analytical and iterative model that takes into account the effect of charge (during the spike) and discharge (between two successive spikes) of the NPs on the drain current as detailed elsewhere.[7] Roughly, the RC charge/discharge time constant of the NOMFET is governed by the channel resistance R of the NOMFET, which scales as L/µ and the capacitance of the NPs network, N·$C_{NP}$, with N the number of NPs, $C_{NP}$ the self-capacitance of a NP $C_{NP} = 2\pi\varepsilon D$ (with D the NP diameter and ε the dielectric constant). Here, at constant L, N and D, the decrease of NOMFET response time constant by a factor 10 is consistent with the increase of mobility by around the same factor (best mobility ≈$10^{-2}$ cm²/Vs in our previously fabricated NOMFET).[7, 8] A more precise comparison is difficult because of device-to-device dispersion, and the density of NP can only be roughly controlled around 5x$10^{10}$ NP/cm² (Fig. S1) as it may vary from device to device by about a factor 2.

In term of energy, the energy per spike $E_{spike}=V_{spike} I_{spike} \delta$, with $V_{spike}$ the applied spike amplitude, $I_{spike}$ the current measured during the spike (Fig. 6) and δ the spike duration, we obtain $E_{spike}$ ~ 2 nJ for the NOMFET with OTS ($V_{spike}$=1V, δ=10ms and taking an average $I_{spike}$ of 200 nA, Fig. 6-a) and $E_{spike}$ ~ 10 nJ for the NOMFET without OTS ($V_{spike}$=10V, δ=100ms and taking an average $I_{spike}$ of 10 nA, Fig. 6-b). Thus, we have gained a factor 5 with the OTS treatment. Note that between spikes, the energy consumption is zero, all the device terminals being grounded.

## 4. Conclusion and perspectives.

In conclusion, we have developed a hybrid metal NP/organic synapstor working at 1 volt and with a typical response time constant in the range 100-200 ms. We have established a relationship between the device performances and i) the presence of Au NPs at the pentacene/ $SiO_2$ interface and ii) the chemical surface treatment before the pentacene deposition. Further reduction of the spike amplitude can be obtained by using high-k dielectric gate or electrolyte gating.[40] Further reduction of the time response will be also possible using recently reported organic semiconductors[41-43] with a higher mobility (> 1 $cm^2/Vs$) then pentacene, or solution-processed (i.e. sol-gel) inorganic oxide semiconductors (up to 7 $cm^2/Vs$).[44]

## Acknowledgments.

This work has been financially supported by the EU 7th framework program [FP7/2007-2013] under grant agreement n° 280772, project "Implantable Organic Nanoelectronics" (iONE-FP7), and partly by agreement n° 318597, project "SYMONE" (Synaptic Molecular Networks for Bio-inspired Information Processing), and partly by the French ANR agency, project n° ANR 12 BS03 010 01 "SYNAPTOR" (Synapse-like transistor and circuits for neuro-inspired computing architecture).

## Appendix A : Supplementary material.

Supplementary figures (S1-S3) associated with this article can be found, in the online version, at http…….

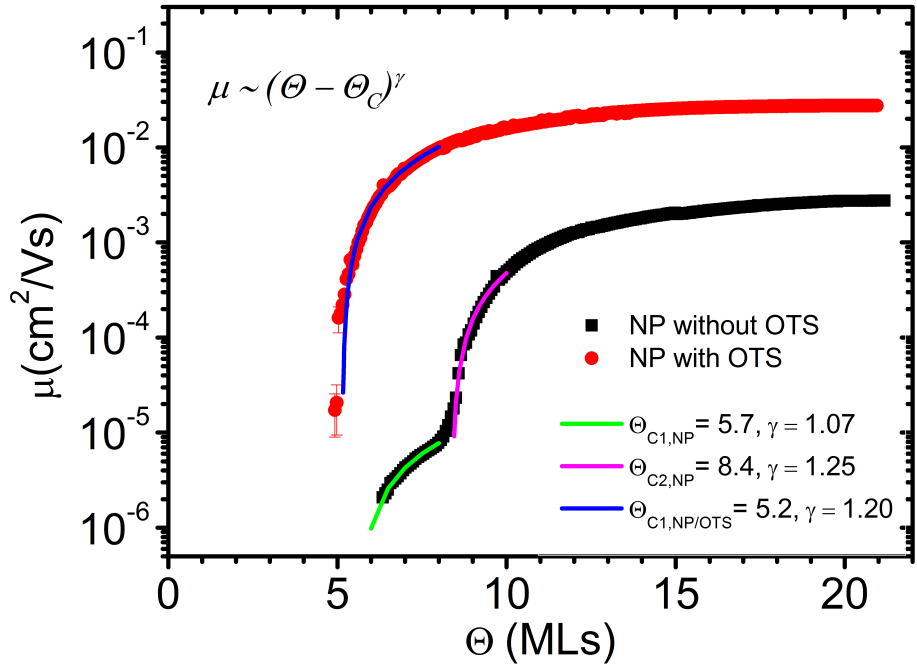

**Fig. 1**. Mobility in saturation µ$_{sat}$ of pentacene NOMFETs as a function of nominal layer thickness Θ measured *in-situ* during the deposition of the semiconducting film. Both data sets were obtained on substrates functionalized with APTMS and 10 nm gold nanoparticles. The red curve was measured using a substrate that had been made hydrophobic by a final OTS functionalization.

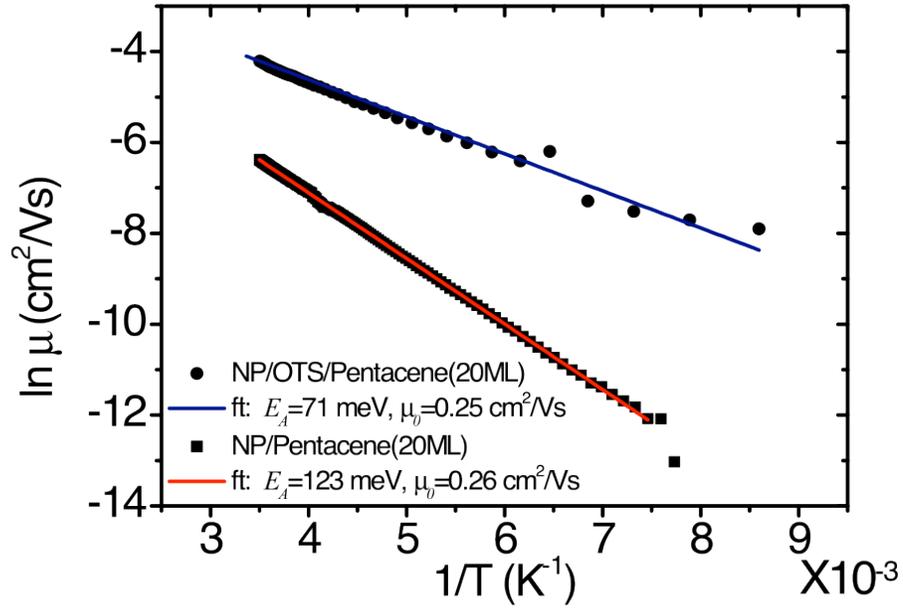

**Fig. 2**. Temperature dependence of mobility in saturation $\mu_{sat}$ for pentacene NOMFETs with 20 ML thickness grown on APTMS/NP (red) or APTMS/NP/OTS surface (blue). Both curves show a temperature activated behavior and follow the Arrhenius behavior. From the slope we determine activation energies as indicated in the plot.

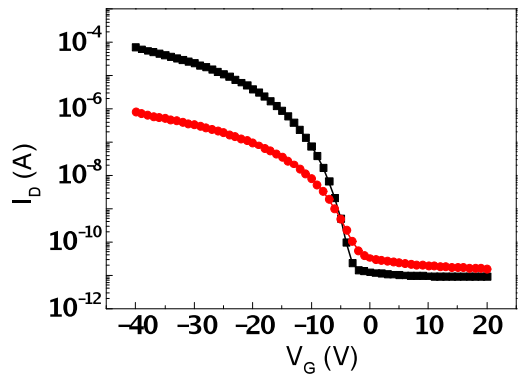

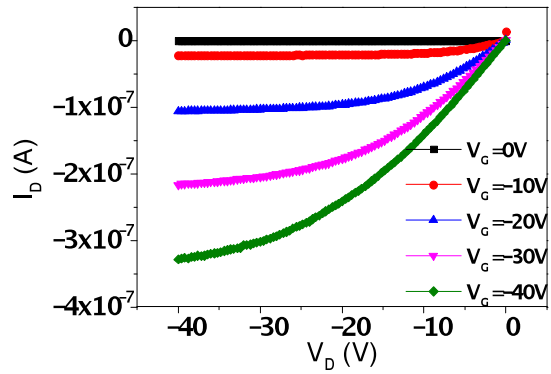

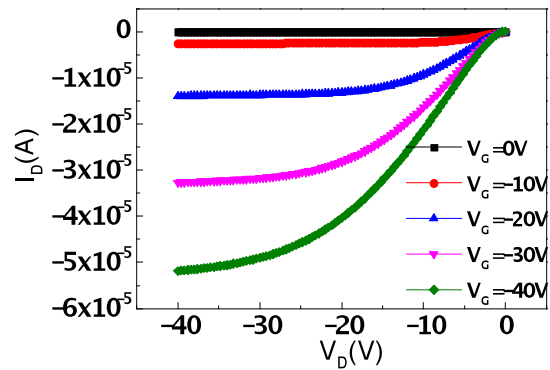

**Fig. 3**. (a) : $I_D$-$V_G$ (at $V_D = -20$ V) for standard NOMFET (APTMS-NP-P5, L=5 μm) without OTS (■) and optimized NOMFET (APTMS-NP-OTS-P5, L=5 μm) with the OTS treatment (●).

$I_D$-$V_D$ characteristics of the same devices : (b) standard NOMFET without OTS and (c) optimized NOMFET with the OTS treatment.

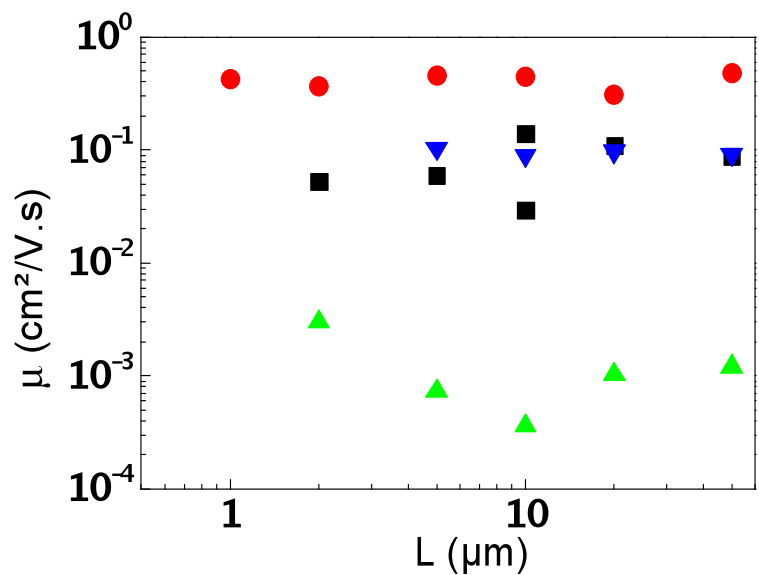

**Fig. 4**. Mobility in saturation for pentacene (P5 ■), pentacene on OTS (OTS-P5 ●) OFETs, for NOMFETs with (APTS-NP-OTS-P5 ▼) and without OTS (APTS-NP-P5 ▲) at different channel length.

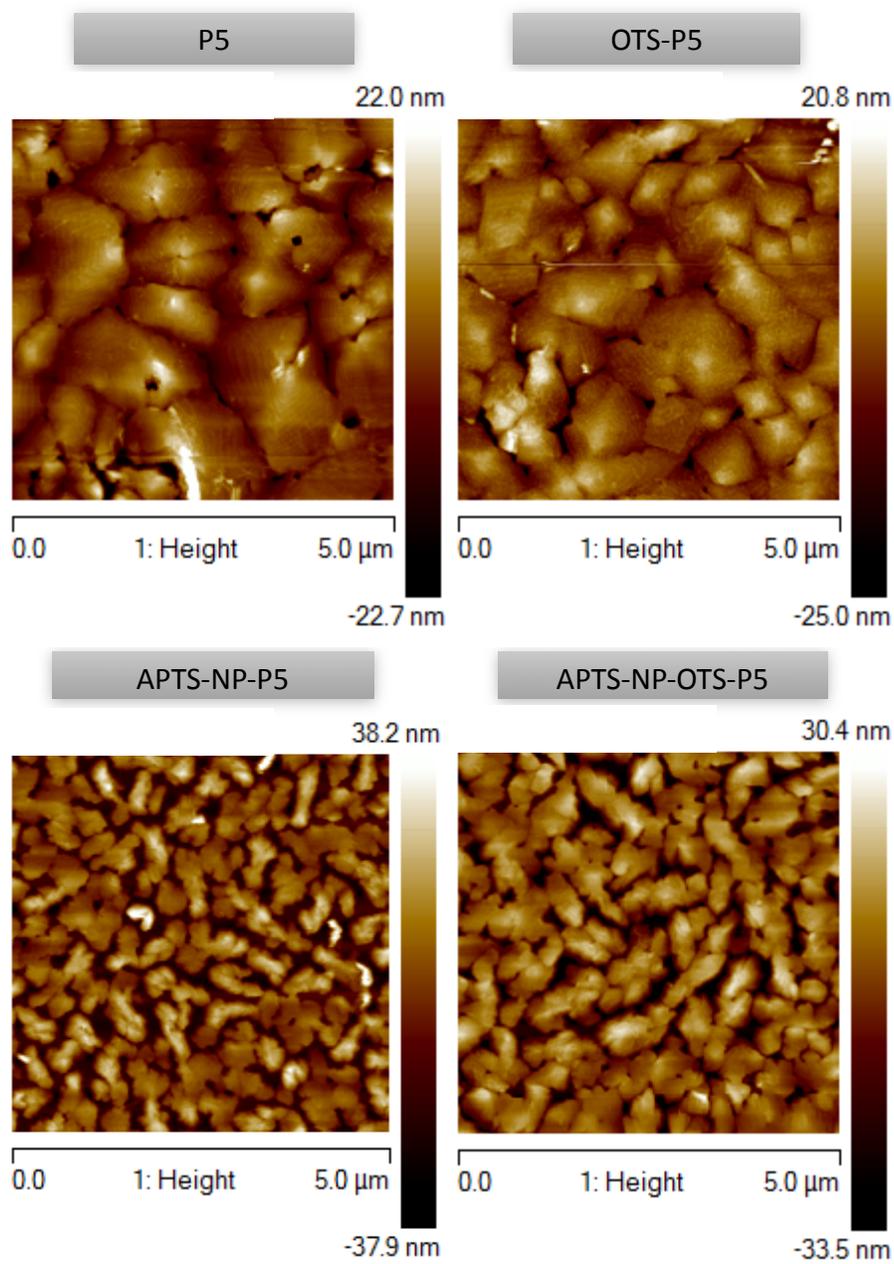

**Fig. 5** Tapping mode-AFM images of the pentacene films for 2 OFETs (top) : pentacene on SiO$_2$ (left) and pentacene on OTS-treated SiO$_2$ (right); and 2 NOMFETs (bottom) : APTMS-NP-P5 (left) and APTMS-NP-OTS-P5 (right) structures. Image analysis (PSD : power spectral density) gives the following average grain size : 1.25 μm, 1.67 μm, 0.55 μm and 0.83 μm, respectively.

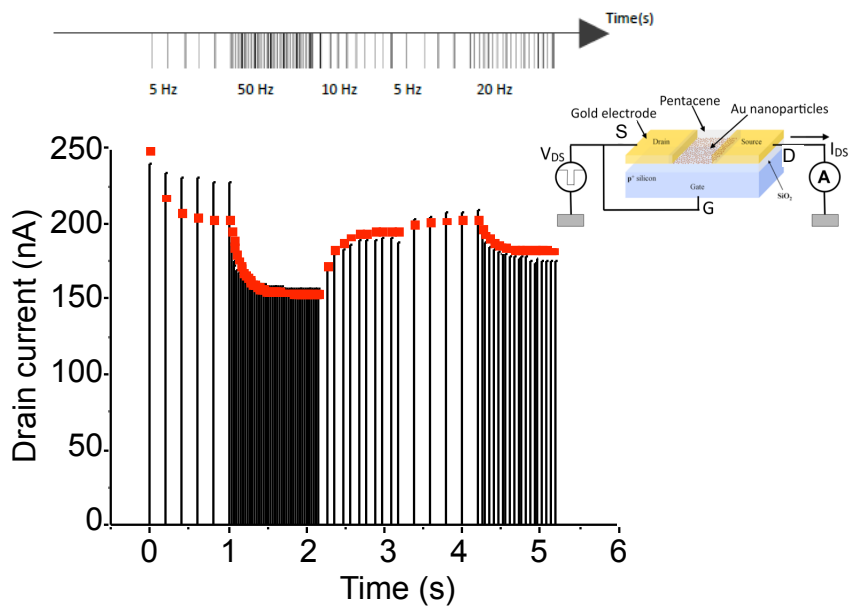

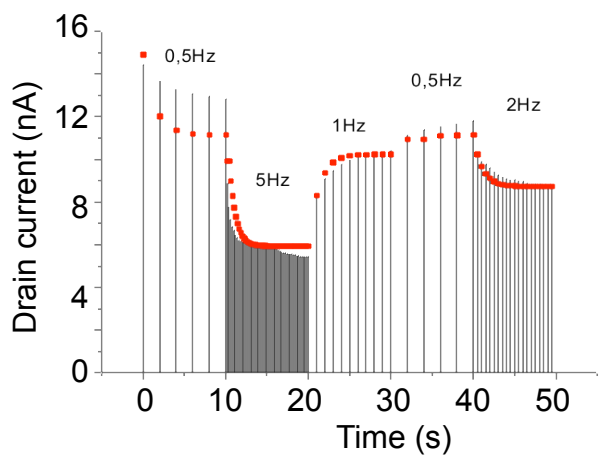

**Fig. 6**. (a) Typical STP response of an optimized NOMFET (L= 1 μm, NPs of 10 nm) subjected to a sequence of spikes at various frequencies (5/50/10/5/20 Hz, see top part of the figure, pulse width = 10 ms) with a pulse amplitude of -1V. The red dots are fits with an analytical model (see Ref. [7]) from which we extract the characteristic response time constant of the NOMFET (here : 187 ms). (b) STP for a standard NOMFET (L= 1 μm, NPs of 10 nm) without the OTS treatment

and for a pulse amplitude of - 10V and spike frequency sequence of 0.5/5/1/0.5/2 Hz (pulse width = 100 ms). The fit gives a STP time constant of 1.5 s.

# SUPPORTING INFORMATION

Fig. S1 shows a SEM image of the NPs on surface before the evaporation of pentacene. From image analysis, an average size of the NPs is 10 ± 3 nm, and a density of about $5\times10^{10}$ NP/cm$^2$.

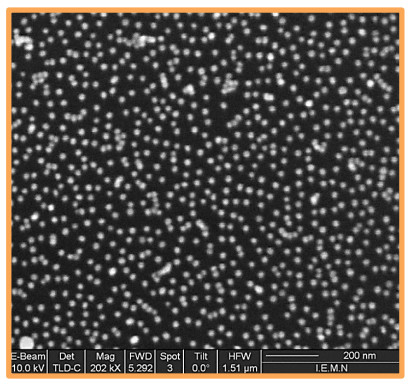

**Fig. S1**. *SEM image (1 μm x 1μm) of the NPs network in the source-drain channel before the pentacene deposition.*

Fig. S2 shows the evolution of the hole mobility in saturation regime as a function of the thickness of the deposited pentacene film. We compare the behavior with and without the NPs. Without the NP, the onset of mobility starts at about 0.8 monolayer (ML) of pentacene in agreement with previous work,[1] corresponding to the onset of the percolation path in 1$^{st}$ pentacene monolayer. The steep increase up to 2ML indicates the formation of the spatially confined channel where field effect induced carriers are generated. Then the slow increase of mobility reaching a plateau of 2-3x10$^{-3}$ cm$^2$/Vs at about 15 ML is the fingerprint of a 2D-3D growth transition.[1, 2]

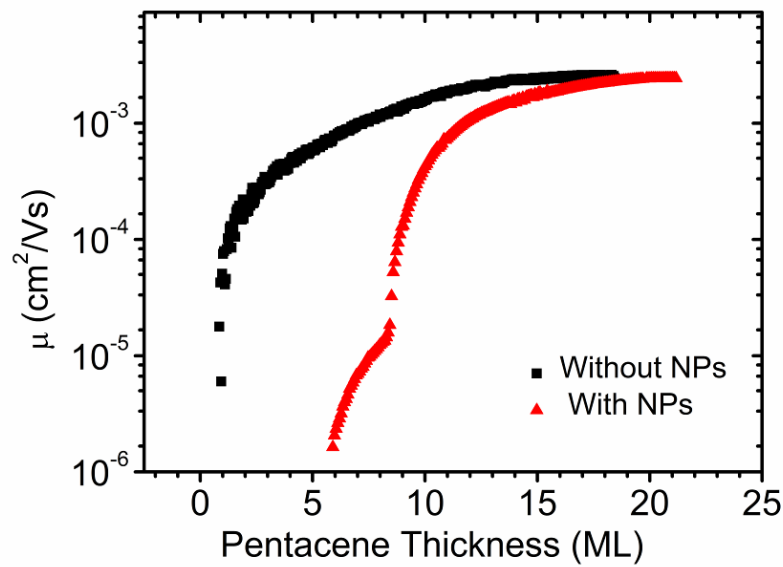

**Fig. S2.** *Evolution of the mobility in saturation versus the pentacene coverage for OFET (without NPs) and NOMFET with NPs (in this latter case, same data as in Fig. 1).*

Figure S3 shows the Arrhenius plot of mobility in saturation measured, *in-situ*, at the end of the pentacene growth, i.e. for a film thickness of about 20 ML (30 nm). The presence of NPs leads to an increase of the activation energy to $E_A$ = 125 meV in contrast to $E_A$ = 65 meV in pure pentacene films. This latter value is in agreement with previous reports for pentacene OFET (in the range 20-80 meV).[3-7]

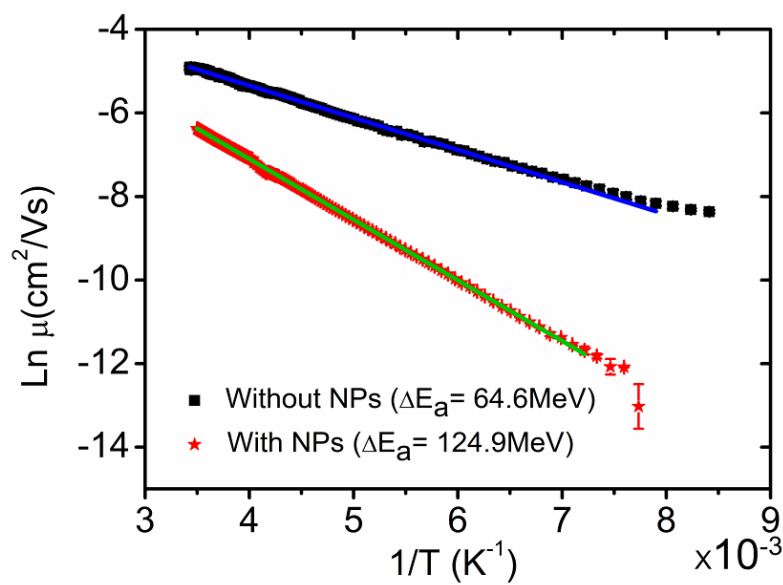

**Fig. S3**. *Arrhenius plot of the temperature dependence of the mobility in saturation for devices with and without NPs. With NPs, data are the same as in Fig. 2 for the APTMS/NP/P5 device.*